\documentclass[aps,prl,twocolumn,showpacs,superscriptaddress,bibliography]{revtex4-1}
\usepackage{amsbsy,amssymb,amsmath,bm}
\usepackage{graphicx}
\usepackage{braket}
\usepackage{float}
\usepackage{textcomp}
\usepackage{color}
\usepackage[colorlinks=true,linkcolor=red]{hyperref}
\usepackage[sort&compress]{natbib}
\usepackage[normalem]{ulem}
\usepackage[shortlabels]{enumitem}
\usepackage[table]{xcolor}
\usepackage{xcolor,colortbl}
\usepackage{soul}
\usepackage{amsmath}
\usepackage{appendix}
\usepackage[english]{babel}

\input{epsf}

\newcommand{\gitrepo}{\url{https://github.com/Kirill-Shulga/Time-Molecule}}

\begin{document}
\title{Dissipative Stability and Dynamical Phase Transition in Two Driven Interacting Qubits}

\author{K. V. Shulga}
\affiliation{International Center for Elementary Particle Physics, The University of Tokyo, 7-3-1 Hongo, Bunkyo, Tokyo 113-0033, Japan}

\date{\today}

\begin{abstract}
We examine a two-qubit system influenced by a time-periodic external field while interacting with a Markovian bath. This scenario significantly impacts the temporal coherence characteristics of the system. By solving the evolution equation for the density matrix operator, we determine the characteristic equilibration time and analyze the concurrence parameter—a key metric for quantifying entanglement. Our findings reveal the system's ability to navigate through a dynamic phase transition. These results pave the way to designing systems of interacting qubits demonstrating robust entanglement under realistic conditions of interaction with the environment.

\end{abstract}

\maketitle

\section{Introduction} 

Quantum computing aims to harness quantum-mechanical phenomena, such as entanglement and superposition of states, to create highly efficient systems of interconnected qubits that could outperform classical computers \cite{ladd2010quantum}. Recent years have witnessed successful implementations of quantum computation principles in systems featuring multiple qubits. Within this context, various studies have advocated using driven qubits as a strategy for stabilization \cite{mohamed2019non,govia2022stabilizing}. The scalability of coherently interacting qubits is pivotal for achieving efficient fault-tolerant quantum computation. Nevertheless, this endeavor presents formidable challenges, including the need to address quantum nonlinearity and coherence suppression or absence, both integral to quantum computation \cite{bernien2017probing, akahoshi2023partially, moody20222022,krasnok2023advancements}.

The interaction of quantum systems with their environment emerges as a primary source of decoherence and dephasing, introducing equilibration and phase-kicks to the system \cite{viola1998dynamical,zurek2003decoherence}. Consequently, dissipation stands out as a formidable obstacle in realizing quantum computing devices. Addressing this challenge has become a current trend, focusing on mitigating the adverse effects of dissipation mechanisms in systems featuring multiple coupled qubits. The complexity inherent in many-body dissipative-driven dynamics poses a significant hurdle. 

This emphasis on mitigating dissipation aligns with the broader exploration of coherent quantum dynamics in systems subject to a time-periodic external perturbation, known as \textit{Floquet quantum systems} \cite{dittrich1998quantum,grifoni1998driven,kohler2005driven}. Extensive attention has been devoted to experimental and theoretical studies in this realm. The investigation of periodically driven quantum systems, composed of many interacting quantum particles, promises a wider variety of temporal interference patterns \cite{khemani2016phase,moessner2017equilibration,d2013many,lazarides2015fate,ponte2015many}. Predictions include the emergence of diverse non-equilibrium coherent quantum phases and quantum phase transitions in spatially extended Floquet quantum systems \cite{khemani2016phase,moessner2017equilibration,sacha2018anderson,chitra2017}. These phases are characterized by a broken temporal-translation symmetry induced by the underlying time-periodic perturbation and exhibit enhanced quantum entanglement.

We focus on a minimal system of two coupled qubits based on a model introduced in \cite{shulga2021time} to establish a foundation for such investigations. This model is designed to elucidate the emergence of phenomena related to temporal coherence in a small number of qubits. Our analysis calculates the equilibration time, highlighting the formation of a dynamic phase transition induced by dissipation. Additionally, we examine the formation and decay of entanglement, employing quantum concurrence as a key metric. This research contributes essential insights to understanding dissipative-driven dynamics in quantum systems, paving the way for future advancements in the design and optimization of quantum computing devices.

\section{Model} 
Let us consider a series of interacting qubits that are subjected to periodic spin-flip pulses controlled by a Floquet Hamiltonian:

\begin{equation} \label{FloquetHamiltonian}
H(t) = 
 \begin{cases}
   \frac{\alpha}{2} \sum_i \sigma_i^x, ~~~~~~~~~~~~~~~~~~~~~0<t<t_1\\
   g\sum_{\langle i,j \rangle} (\sigma_i^x\sigma_j^x + \sigma_i^y\sigma_j^y),~~ t_1<t<T,
 \end{cases}
\end{equation}
where $\sigma_i^{(x,y)}$ are Pauli matrices of the $i$-th qubit, $T$ is the period of Floquet modulation, $g$ is the qubit interaction strength, and $\alpha$ is the normalized amplitude of the periodic flip pulses so $\alpha t_1 = \pi - 2\varepsilon$, where $\varepsilon$ is the detuning from ideal $\pi$-pulse. 

We model dissipative processes using the Lindblad master equation (here and after that $\hbar = 1$):

\begin{equation} \label{Lindbladian}
\frac{d\rho}{dt} = L[\rho] = -i[H(t),\rho] + \sum_k \gamma_k (A_k\rho A_k^\dagger - \frac{1}{2}\{A_k^\dagger A_k, \rho\}), 
\end{equation}
\begin{widetext}
\begin{figure*}
\centering
\includegraphics[width=0.95\textwidth]{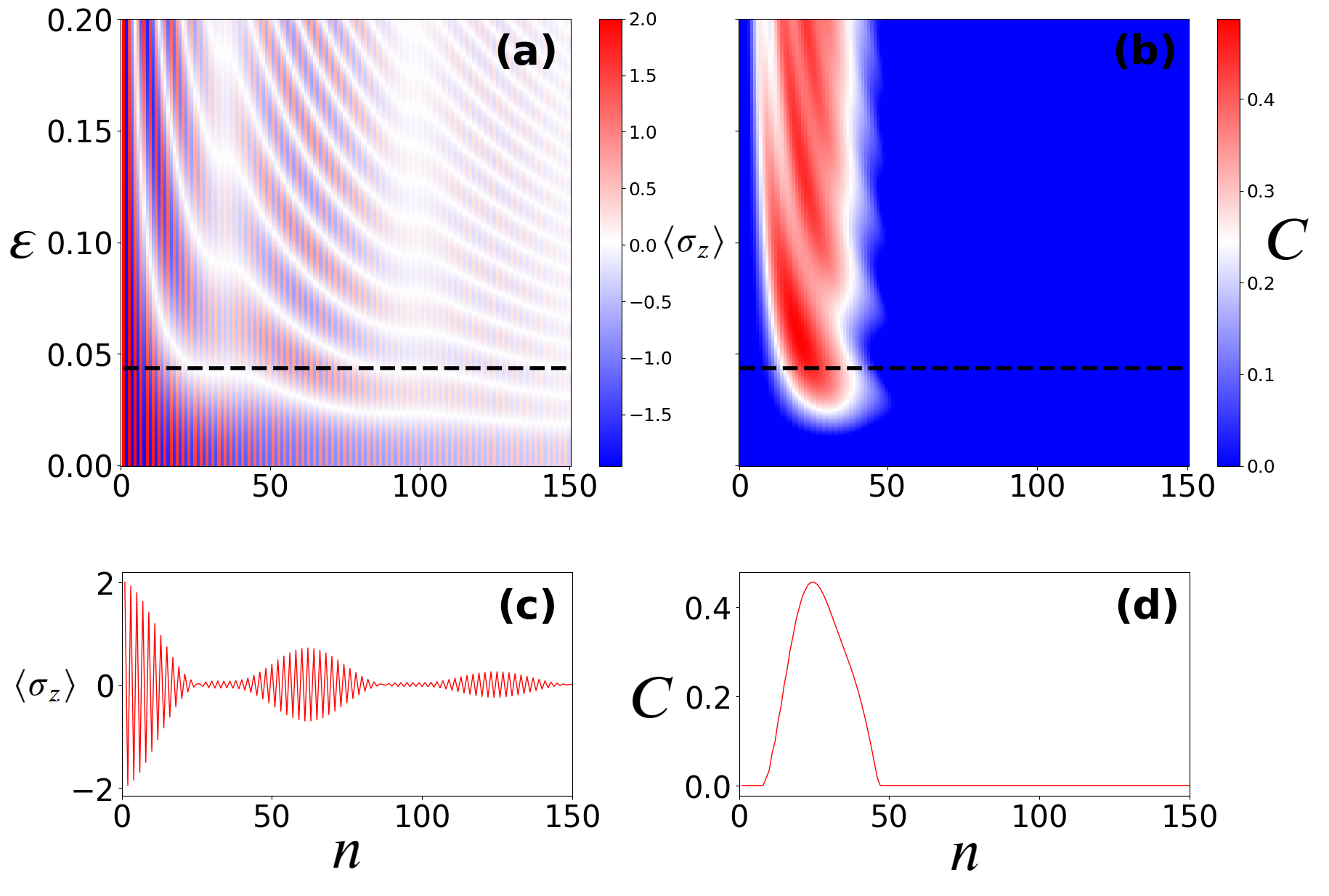}
\caption{ 
(a) Instantaneous total polarization $\langle \hat \sigma_z \rangle$ and (b) concurrence as a function of  the discrete-time $n$ and the pulse imperfection $\varepsilon$ for two interacting qubits. The detuning $\delta_{1,2}$ between the qubits is set to zero. 
(c) and (d) Cross-sections of the plots in (a) and (b) at $\varepsilon=0.0436$ indicated with dashed lines. Except as otherwise noted $g_1~=~0.01$, $g_2~=~0.0025$, $n_{th}~=~0.08$, $T~=~1$, $g~=~0.05.$}
\label{Fig1}
\end{figure*}
\end{widetext}
where $\rho$ is the density matrix of the qubit system, and the sum over dissipation channels $k$ consists of the jump operators $A_k$ with the rates $\gamma_k = \{ \gamma_{i}^+,\gamma_{i}^-,\gamma_{i}^d \}$, where the $i$ index denotes the qubit number. We consider two main sources of decoherence that are typical for qubit experiments. 

The first of them is a coupling to a thermal bosonic bath modeled by the operators $A_i^{\pm} = \sigma_i^{\pm}$  (where $\sigma_i^{\pm} = (\sigma_i^{x} \pm \sigma_i^{y})/2$) with dissipation rates $\gamma_i^- = g_1(1+n_{th})$ and $\gamma_i^+ = g_1n_{th}$, where $g_1$ is the coupling strength between the qubit and the bath, and $n_{th}$ is the thermal distribution of the external bath. The second source of decoherence is the dephasing, modeled by the operator $A_i^d = \sigma^z_i$ with $\gamma_i^d = g_2$. 

When qubits in interaction are coupled to a thermal environment, the system's dynamics are influenced, causing a gradual relaxation toward thermal equilibrium. This, in turn, results in a reduction of the system's entanglement, a phenomenon observable through various parameters including total polarization, entanglement entropy, purity, and concurrence, as illustrated in Fig.\ref{Fig1}.

Various parameters of the system exhibit distinct responses to the presence of coupling with the thermal environment. With an increase in the Floquet cycle number, the system's state becomes more mixed, as evidenced by the shift of the total polarization toward zero, as illustrated in Fig.\ref{Fig1}(a,c). Simultaneously, the entanglement entropy $S$ experiences an increase, signifying the system's enhanced mixing. This augmentation of a mixed state diminishes the quantum correlations between subsystems, reducing information that can be gleaned about the system's state through measurements of individual subsystems. Intriguingly, the quantum concurrence $C$, considered a measure of two-qubit entanglement (see Appendix A), decreases as the system becomes more mixed. Notably, a critical Floquet cycle value exists, beyond which the quantum concurrence reaches zero, and this critical point is proportional to $1/\gamma_k$, as depicted in Fig.\ref{Fig1}(b,d). This paper establishes a foundation for comprehending the dissipative dynamics of phase transitions and the coherence properties inherent in such quantum systems.

\begin{widetext}

\begin{figure}
\centering
\includegraphics[width=\textwidth]{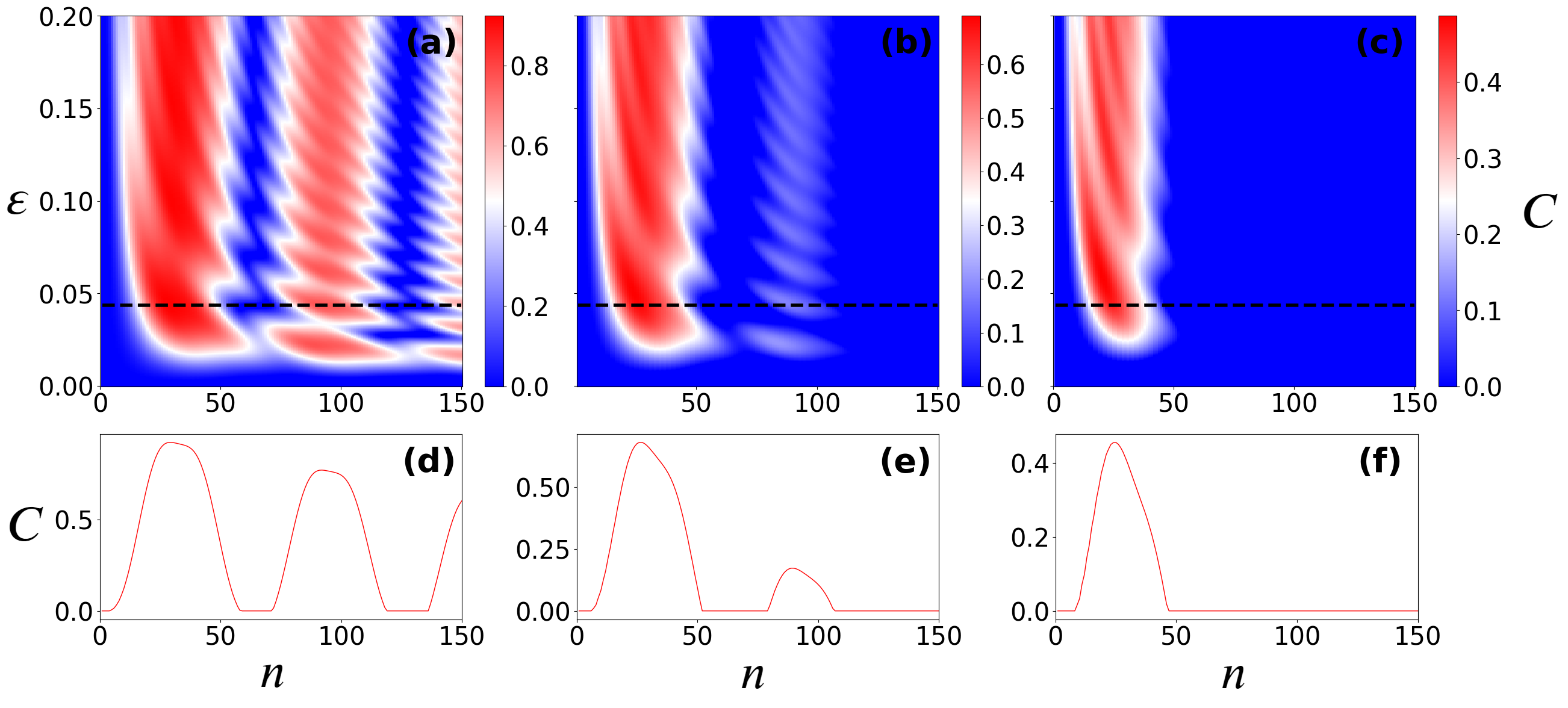}
\caption{ 
(a)--(c) Concurrence as a function of the discrete-time $n$ and the pulse imperfection $\varepsilon$ for two interacting qubits. The detuning $\delta_{1,2}$ between the qubits is set to zero. 
(d)--(f) Cross-sections of the plots in (a)--(c) at $\varepsilon=0.0436$ indicated with dashed lines. Ratio $g_1/g_2=4$, $n_{th}~=~0.08$, $T~=~1$, $g~=~0.05$ for all plots, $g_1~=~0.001$ in (a) and (d), $g_1~=~0.005$ in (b) and (e), $g_1~=~0.01$ in (c) and (f).}
\label{Fig2}
\end{figure}

\end{widetext}

To investigate the variation of concurrence with increasing coupling strength between the two-qubit system and the thermal bath, we select a parameter set closely resembling those achievable in real experiments, with the ratio $g_1/g_2$ fixed (see Appendix B). We assume the system's temperature $n_{th}$ to be lower than the qubit's frequency and explore the persistence and formation of entanglement quantified by concurrence. In Fig.\ref{Fig2}, we illustrate the dynamics of concurrence starting from zero-entanglement initial conditions for different values of $\varepsilon$. Notably, with an increase in the coupling strengths $g_1$ and $g_2$ between the qubits and the thermal bath, the amplitude of the concurrence oscillations decreases. 
The code is available at \footnote[3]{The code is available at \gitrepo.}.

\section{Results} 

Both local observables and concurrence are determined through numerical solutions of (\ref{Lindbladian}). To ascertain the characteristic lifetime of the system, or the \textit{dissipative equilibration time}, we employ a numerical diagonalization approach on the vectorized representation of the Lindbladian superoperator (\ref{Lindbladian}), obtaining the set of eigenvalues $\lambda_i$. The superoperator spectral gap, denoted as $\Delta_1 = -$max(Re$\lambda_i$), is defined as the smallest distance from the imaginary axis to an eigenvalue of $L$, serving as the basis for calculating the equilibration time $t_{eq}=1/\Delta_1$. By equilibration time, we mean that, after initial transients, the expectation value of the quantity in question exhibits slight fluctuations around its average value over an infinite time, thus being very close to this saturation point in most cases. This scale indicates the time during which quantum computations can be performed without occurrence of overwhelming errors.

After a characteristic dissipation time, entanglement vanishes, while oscillations of local observables remain stable and detectable for an extended period, as depicted in Fig.\ref{Fig1}(a,c). This result holds significant implications for quantum computing implementations: even as local time-coherence persists, entanglement may already be lost, rendering computations infeasible. Additionally, we explore the connection between the $t_{eq}$ regimes and concurrence.

\begin{figure}[h!]
\centering
\includegraphics[width=3.4in,angle=0]{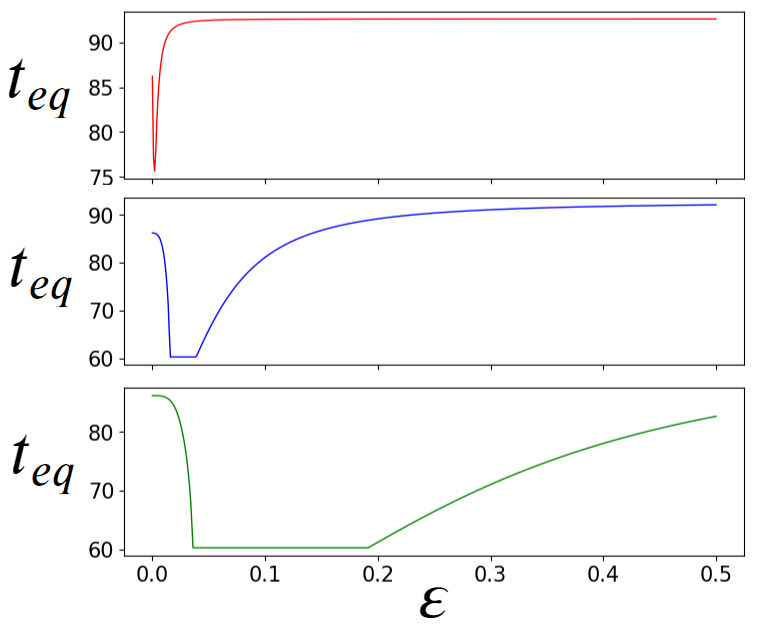}
\caption{Dependence of the equilibration time $t_{eq}$ on detuning $\varepsilon$ at different interaction strength, red $g = 0.001$, blue $g = 0.02$, green $g = 0.1$.}
\label{Fig3}
\end{figure}

The equilibration time, representing the system's operational lifetime is depicted as a function of the detuning parameter in Fig.\ref{Fig3}.
The dependence reveals three distinct regimes: decreasing $t_{eq}$, constant $t_{eq}$, and increasing $t_{eq}$. The emergence of a constant $t_{eq}$ region is linked to the transition from a weakly dissipative regime to the region of large $\varepsilon$. The transition from the decreasing to constant $t_{eq}$ regime exhibits extreme sharpness in the weak dissipation regime. However, the transition from the constant to an increased $t_{eq}$ regime is smoother.

\begin{figure}[h!]
\centering
\includegraphics[width=3.2in,angle=0]{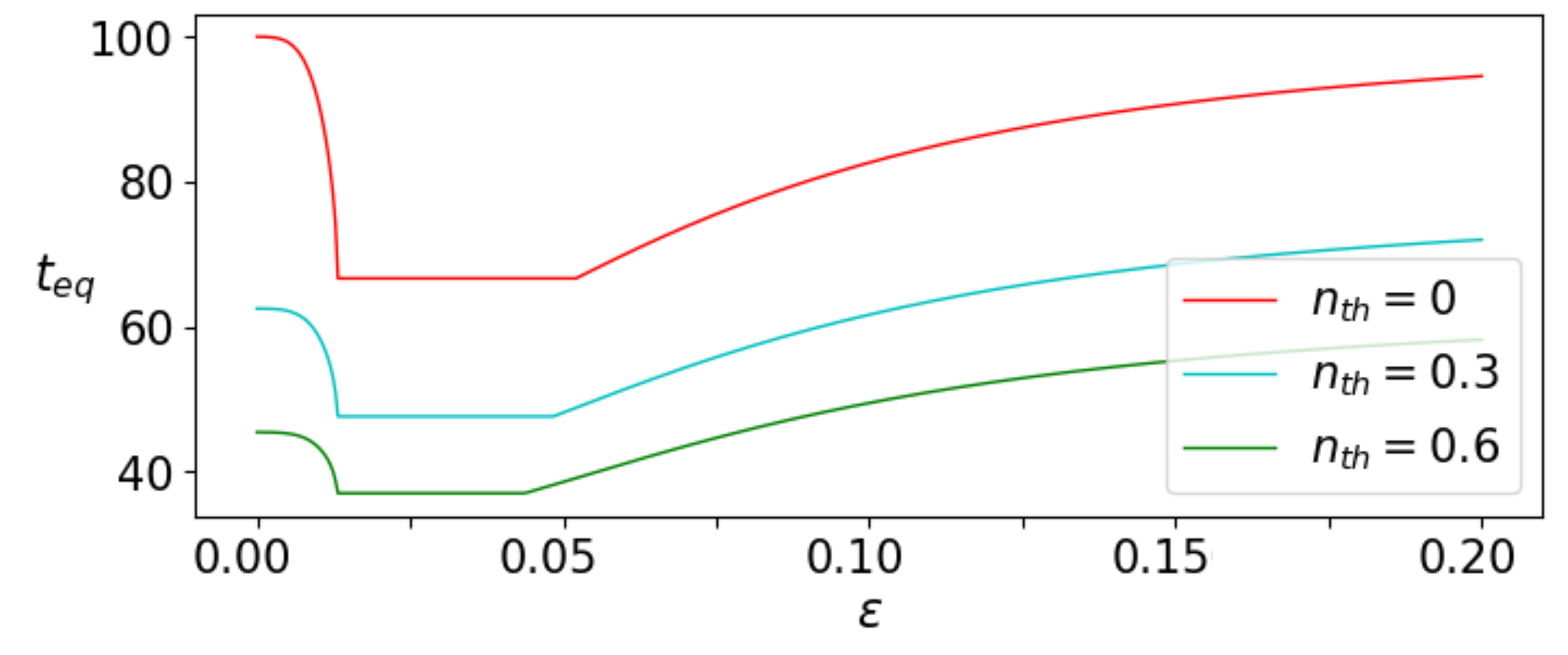}
\caption{Dependence of the equilibration time $t_{eq}$ on detuning $\varepsilon$ at different temperatures of the bath $n_{th}$. Red $n_{th} = 0$, cyan $n_{th} = 0.3$, green $n_{th} = 0.6$. $g_1~=~0.01$, $g_2~=~0.0025$.}
\label{Fig4}
\end{figure}

\begin{figure}[h!]
\centering
\includegraphics[width=3.6in,angle=0]{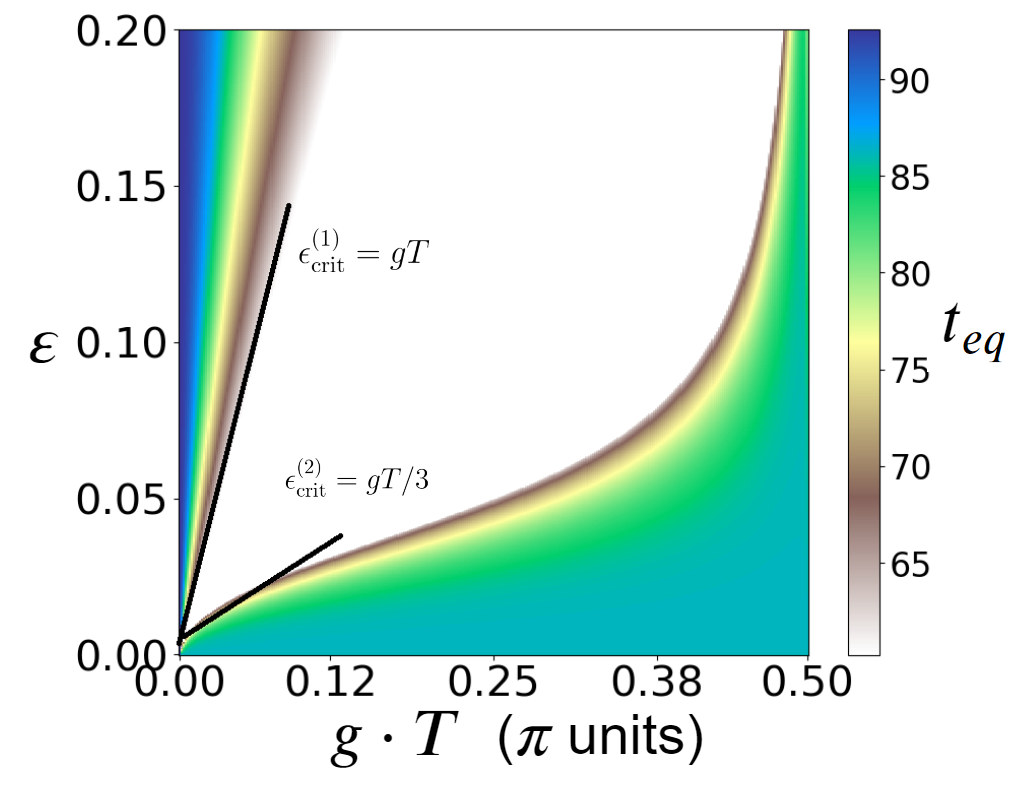}
\caption{Decreasing, stable (white), and increasing lifetime regimes in the control parameter space. Two black lines indicate $\varepsilon = gT/3$, and $\varepsilon = gT$ transition asymptotics.}
\label{Fig5}
\end{figure}

The parameter region characterized by a constant $t_{eq} = const$ proves particularly intriguing for the study and comprehension of dissipative systems. Within this region, the constant equilibration time $t_{eq}$ value diminishes with an increase in bath temperature, as illustrated in Fig.\ref{Fig4}. Moreover, the boundary values of $\varepsilon$, which delineate this region, changes under the influence of temperature (weakly) and the interaction strength $g$ between the two qubits, as shown in Fig.\ref{Fig5}. 

Our observations reveal that the first transition point shifts closer to $\varepsilon = 0$ with decreasing dissipation strength, while the second transition point remains unchanged, maintaining a value of $\varepsilon_{crit} = gT$. 

Since the system operates far from equilibrium, the parametric behavior of the dynamical order parameter $1/t_{eq}$ signifies a dynamical phase transition. As depicted in Fig.\ref{Fig5}, this figure illustrates the three regimes within the complete parametric space, characterized by growth, reduction, or stability against changes in $\varepsilon$. In the weak coupling regime, the transitions follow simple relations, specifically $\varepsilon_{crit}^{(1)} = gT$ and $\varepsilon_{crit}^{(2)} = gT/3$. However, as we move away from this limit, the diagram reveals a significant complexity in dynamics, underscoring the intricate behavior of a system comprised of only two interacting qubits (see Appendix C).

Hence, the upper limit of $\varepsilon_{crit}^{(1)}$ holds limited interest in the context of practical experiments with qubits. This is because a non-ideal control pulse value comparable to the coupling constant is relatively rare. In contrast, the lower limit $\varepsilon_{crit}^{(2)}$ serves as a crucial demarcation between two phases. One phase represents a stable yet short-lived region, wherein fluctuations of parameters exert minimal influence on the final time of quantum entanglement disappearance. The other phase encompasses a region where qubits can sustain entanglement for a more extended period, but external fluctuations have the potential to alter this duration. We posit that the stability exhibited by the region with $t_{eq} = const$ could be leveraged to enhance contemporary quantum algorithms.

\section{Conclusion} 
In conclusion, our study has demonstrated that a dissipative system comprising two qubits under periodic Floquet pumping may exhibit a loss of quantum entanglement, as quantified by concurrence, despite yielding observables indistinguishable from those in the non-dissipative scenario. Furthermore, our analysis of the equilibration time has revealed distinct regions, with particular emphasis on the intriguing $t_{eq} = const$ region. Notably, within this region, the disappearance of coherence persists with an unchanged minimum time, spanning a wide parameter range. This finding underscores the intricate interplay between qubit interactions, Floquet modulations, and dissipation — a phenomenon crucial for various quantum computing applications. Even in the case of just two qubits, the behavior of such systems can manifest complexity. This paper establishes a foundational understanding of the coherence properties and dissipative dynamical phase transitions inherent in these systems, paving the way for further exploration and development in quantum information processing.

\section{Acknowledgments}
We thank Ihor Vakulchyk and Mikhail Fistul for useful discussions and Toshiaki Inada for proofreading the article.

\appendix
\section{Appendix} 

\section{A. Concurrence of two interacting qubits}

Quantum concurrence is a measure of the degree of entanglement between two qubits. Hill and Wootters introduced it \cite{hill1997entanglement} in 1997 as an alternative to the von Neumann entropy of entanglement \cite{neumann1927quantum} for the characterization of the entanglement between two qubits.

The concurrence is defined as follows: consider a two-qubit state represented by the density matrix $\rho$. Then, define the matrix $\tilde{\rho} = (\sigma_y\otimes\sigma_y) \rho^* (\sigma_y\otimes\sigma_y)$, where $\sigma_y$ is the Pauli matrix acting on the y-axis and $\rho^*$ is the complex conjugate of $\rho$ in the standard basis. The concurrence of the state $\rho$ is then given by 
\begin{equation}
C(\rho) = \max(0,\lambda_1 - \lambda_2 - \lambda_3 - \lambda_4),
\end{equation}
where $\lambda_i$ are the eigenvalues of the non-Hermitian matrix $R = \sqrt{\sqrt{\rho}\tilde{\rho}\sqrt{\rho}}$, sorted in decreasing order.

One of the advantages of the concurrence over the von Neumann entropy of entanglement is that it can be extended to mixed states. In particular, for a mixed state $\rho$, the concurrence is defined as the convex roof of the concurrence of the pure state decompositions of $\rho$. This allows for more accurate characterization of the entanglement in realistic scenarios, where the system's state is typically mixed.

Moreover, the concurrence has a simple geometric interpretation: it is related to the distance of the state $\rho$ to the set of separable states. In particular, a state is separable if and only if its concurrence is zero. Therefore, the concurrence provides a clear criterion for identifying entangled states.

\section{B. Selecting the ratio between $g_1$ and $g_2$}

The Kramers-Kronig \cite{kronig1926theory, kramers1927diffusion} relation is a fundamental result in optics, condensed matter physics, and other fields, connecting the real and imaginary parts of the complex response function. In the context of quantum information, the Kramers-Kronig relation is often invoked in the analysis of qubits relaxation and dephasing times, which are typically characterized by the parameters $T_1$ and $T_2$, respectively. The Kramers-Kronig relation provides an upper bound on $T_1$ in terms of $T_2$, given by $T_1 \leq 2T_2$.

Interestingly, this inequality becomes an equality in the limit where the spectral density of the noise is a Lorentzian function (Kubo's formula \cite{kubo1957statistical}). In this case, the dynamics of the qubit can be described by the Bloch-Redfield master equation \cite{cohen1998atom,breuer2002theory}, which contains two coupling constants, $g_1$ and $g_2$, corresponding to the relaxation and dephasing processes, respectively. In the Kramers-Kronig limit, these coupling constants become related by $g_1 \ge \sqrt{2}g_2$. In our paper, we work with a slightly larger relaxation rate that satisfies this inequality, which is much more common in experiments with real qubits.

This result has important implications for engineering qubits with long coherence times. In particular, it suggests that reducing the dephasing rate $g_2$ may not necessarily improve $T_1$, and vice versa. Rather, a delicate balance between these two processes must be maintained to achieve optimal qubit performance. Understanding this balance is crucial for designing and operating future quantum devices, and the Kramers-Kronig relation provides a powerful tool for analyzing and optimizing their performance.

\section{C. Complex behavior of different equilibration time regions}

\begin{figure}[h!]
\centering
\includegraphics[width=3.4in,angle=0]{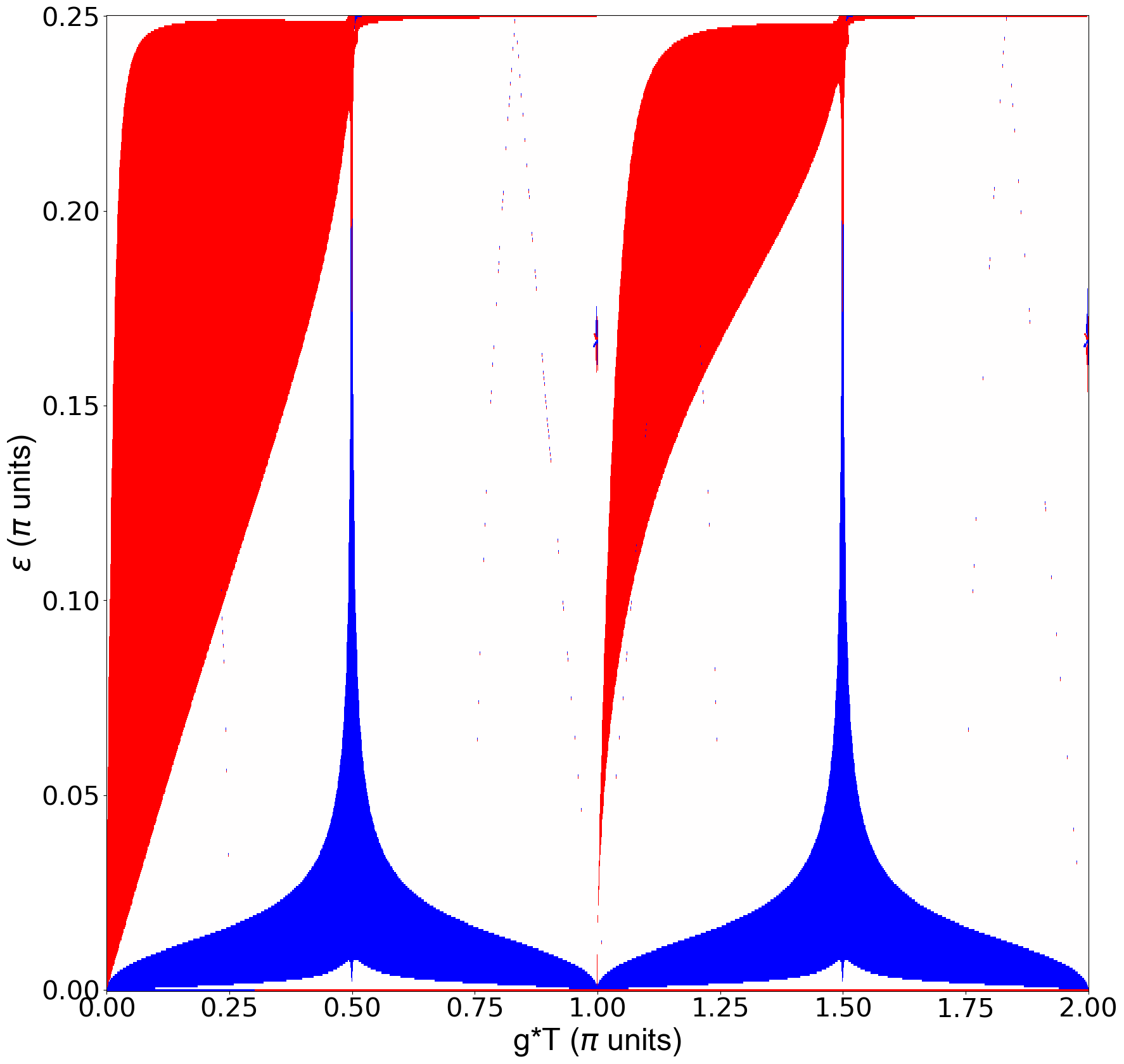}
\caption{Diagram of equilibration time regions in a wide range of epsilon and g parameters. Three regions are highlighted, increasing (red), decreasing (blue) and constant (white) time. Periodicity is visible, as well as artifacts appearing in the form of broken dotted lines in the area $t_{eq}$.}
\label{Fig7}
\end{figure}

We explored the positions of three distinct regions concerning the behavior of equilibration time in the parameter space of $\varepsilon$ and $g$ over a wide range. This dependence exhibits periodic behavior, with discernible shifts in the system's dynamics at very large coupling values, see Fig.\ref{Fig7}. Within the $t_{eq} = const$ region, the emergence of slender dotted lines is notable, and the intermittency of these lines intensifies with decreasing bath temperature. These lines are double, representing a sudden increase and subsequent return to the initial constant value of equilibration time within a very narrow range of parameters. Intriguingly, these artifacts vanish entirely at zero temperature. The mechanism underlying the appearance of additional segments of such lines remains an interesting puzzle for us, warranting further investigation and exploration.

~~

\bibliography{main}

\end{document}